\documentclass[10pt, showpacs, preprintnumbers, nofootinbib, amsmath, amssymb, aps, prc, twocolumn, groupedaddress, superscriptaddress, showkeys]{revtex4-2}
\usepackage{graphicx}
\usepackage{dcolumn}
\usepackage{bm}
\usepackage[colorlinks=true,urlcolor=blue,citecolor=blue]{hyperref}
\usepackage{color,lineno}
\usepackage{booktabs}
\usepackage{mathtools}
\usepackage{amsmath}
\usepackage{subfig}
\usepackage{natbib}
\usepackage{caption}
\begin{document}

\title{Electromagnetic Properties of the N=50 Isotones with the p35-i3 Hamiltonian}

\author{J. A. Purcell and B. A. Brown}
\affiliation{Department of Physics and Astronomy, Michigan State University, East Lansing, Michigan 
48824-1321,
USA
and Facility for Rare Isotope Beams, Michigan State University, East Lansing, Michigan 48824-1321,
USA}

\begin{abstract}
The nuclei with 50 neutrons that lie between $^{78}$Ni and $^{100}$Sn have provided benchmark studies of the nuclear shell model 
for protons in the  $\{0f_{5/2}, 1p_{3/2}, 1p_{1/2}, 0g_{9/2}\}$ model space. 
New Hamiltonians for this model space have recently been 
obtained based on valence-space in-medium renormalization-group (VS-IMSRG) methods 
with two- and three-nucleon interactions.
The two-body matrix elements (TBME) obtained from these ab-initio methods 
served as the starting point for singular-value decomposition (SVD) method fits
of the TBME to experimental binding energies and excitation energies, resulting
in Hamiltonians called p35-i2, p35-i3 and p30-i3.
In this paper,  magnetic moments, quadrupole moments, $B(M1)$ and $B(E2)$ values
obtained with these Hamiltonians are presented and compared to experiment.
Results obtained from various Hamiltonians are compared to assess the theoretical uncertainties.

\end{abstract}
\maketitle


\section{Introduction}

The properties of isotones with $  N=50  $ have historically been a 
testing ground for shell-model configuration interaction (CI) calculations. 
In this paper, We consider electromagnetic data for low-lying states of nuclei between $^{78}$Ni and $^{100}$Sn that can be described by protons in 
the $  \left\{0f_{5/2}, 1p_{3/2}, 1p_{1/2}, 0g_{9/2}\right\} $ $ (\pi j4  $) model space. The region between $^{90}$Zr and $^{100}$Sn 
is well established territory where many wavefunctions are dominated by $  \left\{1p_{1/2}, 0g_{9/2}\right\}  $ 
orbitals \cite{n50Talmi, cohen64, aurbach65, vervier66, ball72, gloeckner73, blomqvist85}. The full $  \pi j4  $  model space has previously been considered with Hamiltonians derived with the 
singular-value decomposition method (SVD) \cite{jw88em, jj44a}, 
as well as those obtained valence-space in-medium renormalization-group (VS-IMSRG) methods \cite{Yuan24}. This paper is based on a set of newly derived SVD Hamiltonians called p30-i3, p35-i2, and p35-i3 \cite{p35i3}. In the notation px-iy, x stands for the number of SVD parameters and y stands for the ab initio starting points used in \cite{p35i3}.

The starting points for the Hamiltonians derived in \cite{p35i3} are those obtained with the VS-IMSRG method. 
These are obtained using the EM 1.8/2.0 NN$+$3N interaction \cite{Heb2011} 
in a harmonic oscillator basis with frequency $  \hbar \omega=12  $ MeV, truncated to 13 major shells ($  2n+l \leq  $ $  e_{\rm max}=12  $). 
The basis is normal ordered with respect to the Hartree-Fock ground state of the reference and the residual 3N interaction was discarded. 
The $  \pi j4  $ valence space was then decoupled using the Magnus formulation of the IMSRG. 
The results obtained with the standard approximation \cite{VSIMSRG}, truncating all operators at the two-body level throughout the flow, 
including inside nested commutators, are labeled IMSRG(2) in \cite{p35i3}. 
The Hamiltonians obtained with the  singular-value decomposition (SVD) method starting with IMSRG(2) are labeled $  i2  $.

Another starting point labeled IMSRG(3f2) in \cite{p35i3}
introduces a correction in which intermediate three-body operators that arise in nested commutators are incorporated by rewriting the double
commutator in a factorized form while maintaining the same computational
scaling as the IMSRG(2) approximation \cite{He24}. As in ref. \cite{He24}, 
factorized terms with one-body intermediate states during the flow
and two-body intermediate states at the end of the flow were included. 
The Hamiltonians obtained with the SVD method starting with IMSRG(3f2) are are labeled $i3$.

These procedure were carried out for two different references states, $^{78}$Ni and $^{100}$Sn, corresponding to empty and full valence spaces, 
respectively, The average of these provided the 65 two-body matrix elements (TBME) for the starting points for the SVD fits. 

It was shown in \cite{p35i3} that the IMSRG(3f2) input provided a better starting point for 
describing the experimental energy data in the sense that the rms deviation obtained with $p<10$ SVD parameters 
was much smaller with IMSRG(3f2) compared to IMSRG(2). Thus, IMSRG(3f2) might provide a better 
input for the linear combinations of SVD parameters that are not well determined from experimental data.

Based on the minimum in the RMS deviations between the Hamiltonian predictions 
and the experimental validation data, 
the optimal number of SVD parameters
was $p=35$ \cite{p35i3}. 
For the purpose of investigating the Hamiltonian uncertainties in the observables, 
we also obtained a Hamiltonian with $p=30$ varied SVD parameters.


\section{Electromagnetic observables}
One can express the one-body reduced matrix element for the $  n  $-particle wave function in the form of a product over one-body transition densities (OBTD) times reduced single-particle matrix elements (SPME) $  \left\langle k_{\alpha }\left \Vert O^{\lambda } \right \Vert k_{\beta }\right\rangle  $:
\begin{equation}
    \begin{aligned}
        \left \langle f\left\Vert\hat{O}^{\lambda }\right\Vert i \right \rangle &=  
        \left \langle n \omega _{f} J_{f}\left\Vert\hat{O}^{\lambda }\right\Vert n \omega _{i} J_{i}\right\rangle \\ &=
        \sum_{k_{\alpha }, k_{\beta }}{\rm OBTD}(f i k_{\alpha } k_{\beta } \lambda )
        \left\langle k_{\alpha }\left\Vert O^{\lambda }\right\Vert k_{\beta }\right\rangle,
    \end{aligned}
    \label{eq:reducedME}
\end{equation}
where the OBTD are given by
\begin{equation}
    {\rm OBTD}(f i k_{\alpha } k_{\beta } \lambda ) = 
    \frac{\left\langle n \omega _{f} J_{f}\left\Vert \left[a^{+}_{k_{\alpha }}\otimes \tilde{a}_{k_{\beta }}\right]^{\lambda }\right\Vert n \omega 
    _{i} J_{i}\right\rangle}
    {\sqrt{ (2\lambda +1)}}
    \label{eq:OBTD}
\end{equation}

The labels $  i  $ and $  f  $ are a shorthand notation for the initial- and final-state quantum numbers 
$  (n \omega _{f} J_{f})  $ and $  (n \omega _{i} J_{i})  $, respectively. The label $  k_{\alpha }  $ indicates 
the spherical single-particle states $  (n,\ell ,j)  $ used in the model space. The NuShellX code \cite{nushellx} 
provides the wavefunctions and the resulting OBTD.
The $  \left\langle k_{\alpha }\left\Vert O^{\lambda }\right\Vert k_{\beta }\right\rangle  $ are the SPME
obtained with the $  E\lambda   $ and $  M\lambda   $ operators.
The standard forms of the SPME are given in \cite{jw88em}.

We consider magnetic moments
\begin{equation}
    \begin{aligned}
        \mu  &= \sqrt{\frac{4\pi }{3}}   \left\langle i,J,M=J \left\vert {\hat O}(M1)\right\vert i,J,M=J\right\rangle \\ &=
        \sqrt{\frac{4\pi}{3}} \left( \begin{array}{ccc} J & 1 & J \\ -J & 0 & J \end{array}\right)
        \left\langle i,J\left\Vert {\hat O}(M1)\right\Vert i,J\right\rangle,
    \end{aligned}
    \label{eq:magnetic}
\end{equation}
quadrupole moments
\begin{equation}
    \begin{aligned}
        Q &= \sqrt{\frac{16\pi }{5}} \left\langle J,M=J\left\vert{\hat O}(E2)\right\vert J,M=J\right\rangle \\ &=
        \sqrt{\frac{16\pi }{5}} {\left( \begin{array}{ccc} J & 2 & J \\ -J & 0 & J \end{array}\right) } \left\langle i,J\left\Vert {\hat O}(E2)\right\Vert i,J\right\rangle,
    \end{aligned}
    \label{eq:quadrupole}
\end{equation}
and reduced transition probabilities
\begin{equation}
    B\left(O^{\lambda },i \rightarrow f\right) = 
    \frac{\left\vert \left\langle f\left\Vert \hat{O}^{\lambda }\right\Vert i \right\rangle \right\vert ^{2}}{2J_{i}+1}    
    \label{eq:reducedProb}
\end{equation}
We consider results for $B(M1)$ and $B(E2)$.
$B(E1)$ are zero 
since the parity-changing $\lambda$=1 operator does not connect any of the orbitals in the $  j4  $ model space. $B(M2)$ are not 
considered since the parity-changing $\lambda$=2 operator is limited to only one ($  0g_{9/2}-1f_{5/2}  $) of the 10  SPME, with one 
orbital in the $  j4  $ model space involved in M2 transitions. 
$B(E3)$ obtained from our wavefunctions are discussed in \cite{be3}.

For $  M1  $  the SPME involve spin and orbital g-factors which are treated as effective parameters 
to take into account configuration mixing outside the $  j44  $ model space and mesonic-exchange currents. 
We use the orbital-dependent effective g-factors obtained from fits to the data in \cite{jw88em}. 

For $  E2  $ the SPME involve radial integrals. It is common to use harmonic-oscillator radial wavefunctions. 
For example,  a value of $  b^{2}=4.481  $ was used in \cite{jw88em} for the oscillator parameter. 
In addition, one uses an effective proton charge $  e_{p}  $ to take into account mixing with configurations that are 
not included in the $  j4  $ model space. In \cite{jw88em} a value of $  e_{p}=2.0  $ was needed with their harmonic-oscillator radial wavefunction assumption. 
Here we  use more realistic nucleus-dependent radial wavefunctions obtained with energy-density functionals (EDF). 
For this purpose, we use the Skx Skyrme functional \cite{skx}. With this EDF, a value of $  e_{p}=1.8  $ gives 
a good reproduction of all the $  E2  $ data we consider (see Sec. IV).


\section{Uncertainties related to the choice of Hamiltonian}

In this section we compare global sets of electromagnetic observables obtained with 
the new p35-i2, p35-i3, and p30-i3 Hamiltonians together the older Hamiltonians jj44a \cite{jj44a}. 
The purpose is to use these comparisons to deduce Hamiltonian uncertainties in the observables. 
All of these Hamiltonians are based on fits to data in the $  \pi j4  $ model space.

We calculated observables for all nuclei between $^{79}$Cu and $^{99}$In, and divided the 
results into those for nuclei with $  A < 88  $ and those for nuclei with $  A \geq 88  $. 
The wavefunctions for low-lying states with $  A<88  $ are dominated by the three orbitals $  \{0f_{5/2},1p_{3/2},1p_{1/2}\}  $ 
and those with $  A \geq 88  $ are dominated by the two orbitals $  \{1p_{1/2},0g_{9/2}\}  $. 
In \cite{p35i3} it was found that the RMS deviation between experiment and theory for the binding energies 
and excitation energies was significantly larger in the first group (0.12 MeV) than in the second group (0.05 MeV). 
Thus, we can anticipate that the RMS deviations in the electromagnetic observables will be larger in the first group compared to those in the second group.

\begin{figure}
\includegraphics[scale=0.40]{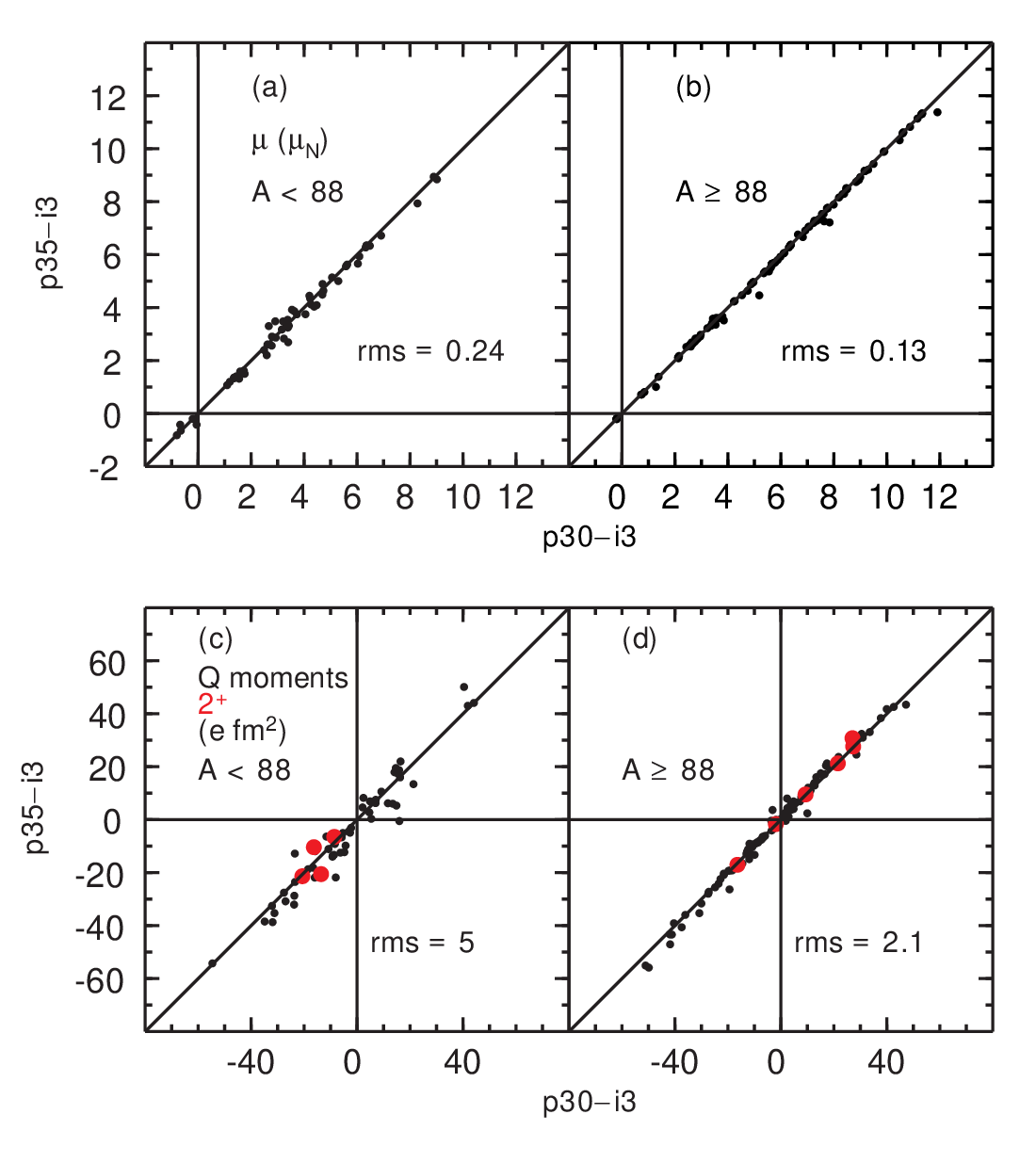}
\caption{Comparison of  moments obtained with the
p30-i3 and p35-i3 Hamiltonians.
}
\label{(1)}
\end{figure}

\begin{figure}
\includegraphics[scale=0.40]{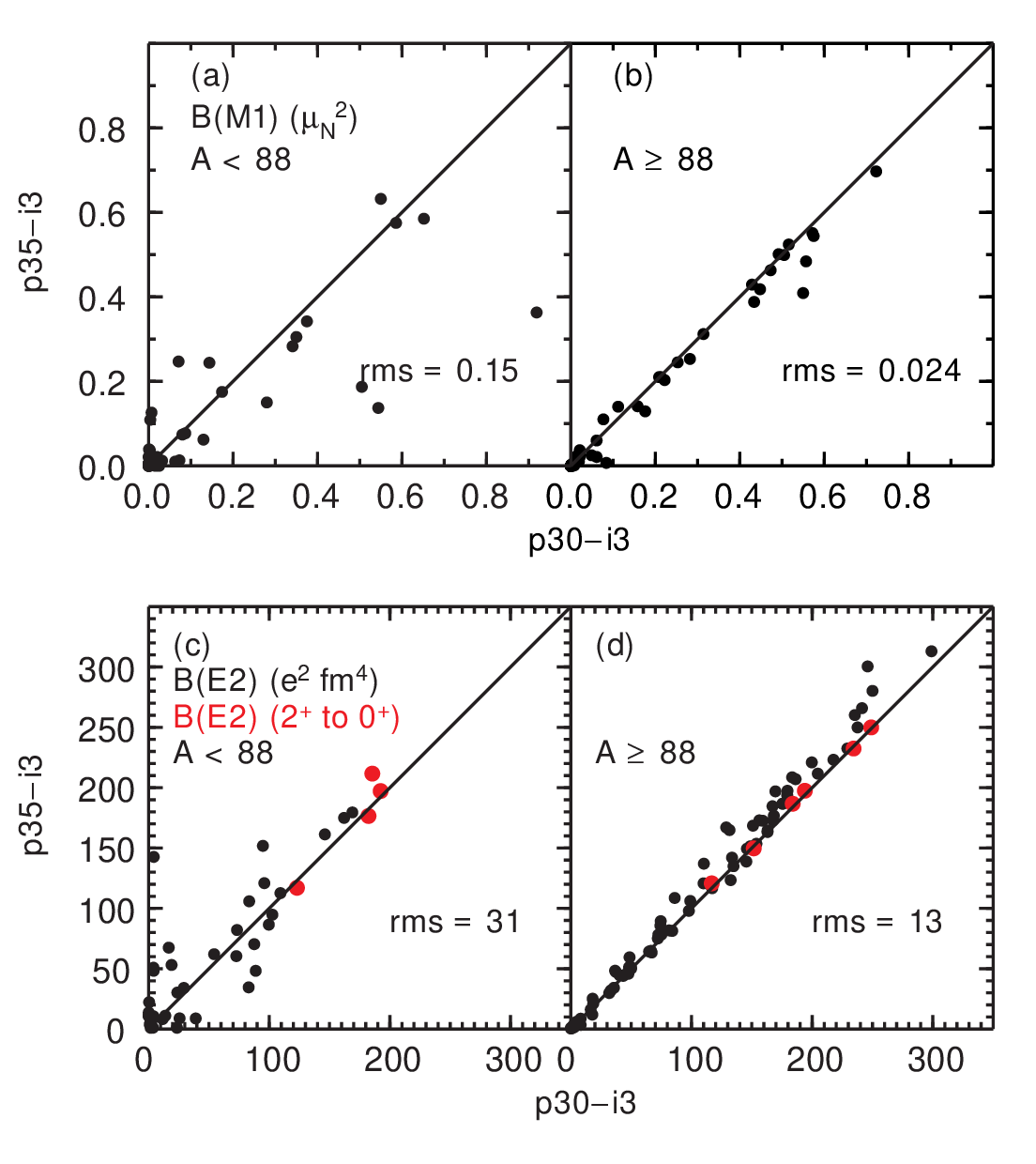}
\caption{Comparison of $B(M1)$ and $B(E2)$ values obtained with the
p30-i3 and p35-i3 Hamiltonians.
}
\label{(2)}
\end{figure}

For each Hamiltonian, we calculated the magnetic and quadrupole moments for the first state of each $  J^{\pi }  $. 
We also calculated the $B(M1)$ and $B(E2)$ values for the first state of each $  J^{\pi }  $ with the first state with $  (J+2)^{\pi }  $. 
We did not include those for which the states were less than 200 keV from the second state with the same $  J^{\pi }  $. 
For each set of comparisons, we give the RMS differences for the specific observables. These values are indicative of the associated theoretical uncertainties.

in Figs. \ref{(1)} and \ref{(2)}, we show the comparison of results obtained with the starting Hamiltonian IMSRG(3f2) for two 
different values of the number of VLC linear combinations varied (30 and 35). 
As expected, the RMS difference is larger for nuclei with $  A<88  $ compared to those with $  A \geq 88  $. 
For the moments (Fig. \ref{(1)}) and $B(E2)$ (Fig. \ref{(2)}-cd), this difference is about a factor of two. 
For the $B(M1)$ values, the rms differences are relatively larger. 
The reason is that the small $B(M1)$ (WU = 1.79 $\mu_{N}^{2}$) arises from the cancellations 
between components in Eq. \ref{eq:reducedME} which  are particularly sensitive to changes in the Hamiltonian.

We find a particular outlier for the $B(E2)$ in Fig. \ref{(2)}c with 
(p30-i3, p35-i3) values of (4, 142) e$^{2}$ fm$^{4}$, that comes from the 7/2$_{1}^{-}$ to 3/2$_{1}^{-}$ 
transition in $^{81}$Ga. As seen in Fig. 9 of \cite{p35i3}, there are two low-lying 3/2$^{-}$ states in $^{81}$Ga. 
The $B(E2)$ for the 7/2$_{1}^{-}$ to 3/2$_{2}^{-}$ transition are (74, 8) e$^{2}$ fm$^{4}$. 
It is not clear what part of the Hamiltonian is responsible for the interchange of properties for these $B(E2)$ transitions.

The calculated gamma-decay branching of the 7/2$^{-}$ state to the (3/2$^{-}_{1}$, 5/2$^{-}_{1}$) 
states is (25, 75)\% for p35-i3 and (1, 99)\% for p30-i3. The gamma-decay data of \cite{ga81} give (0, 100)\% 
for the state at 1.40 MeV. Assuming this is the 7/2$^{-}$ state, the data agree with the p30-i3 Hamiltonian result.

\begin{figure}
\includegraphics[scale=0.40]{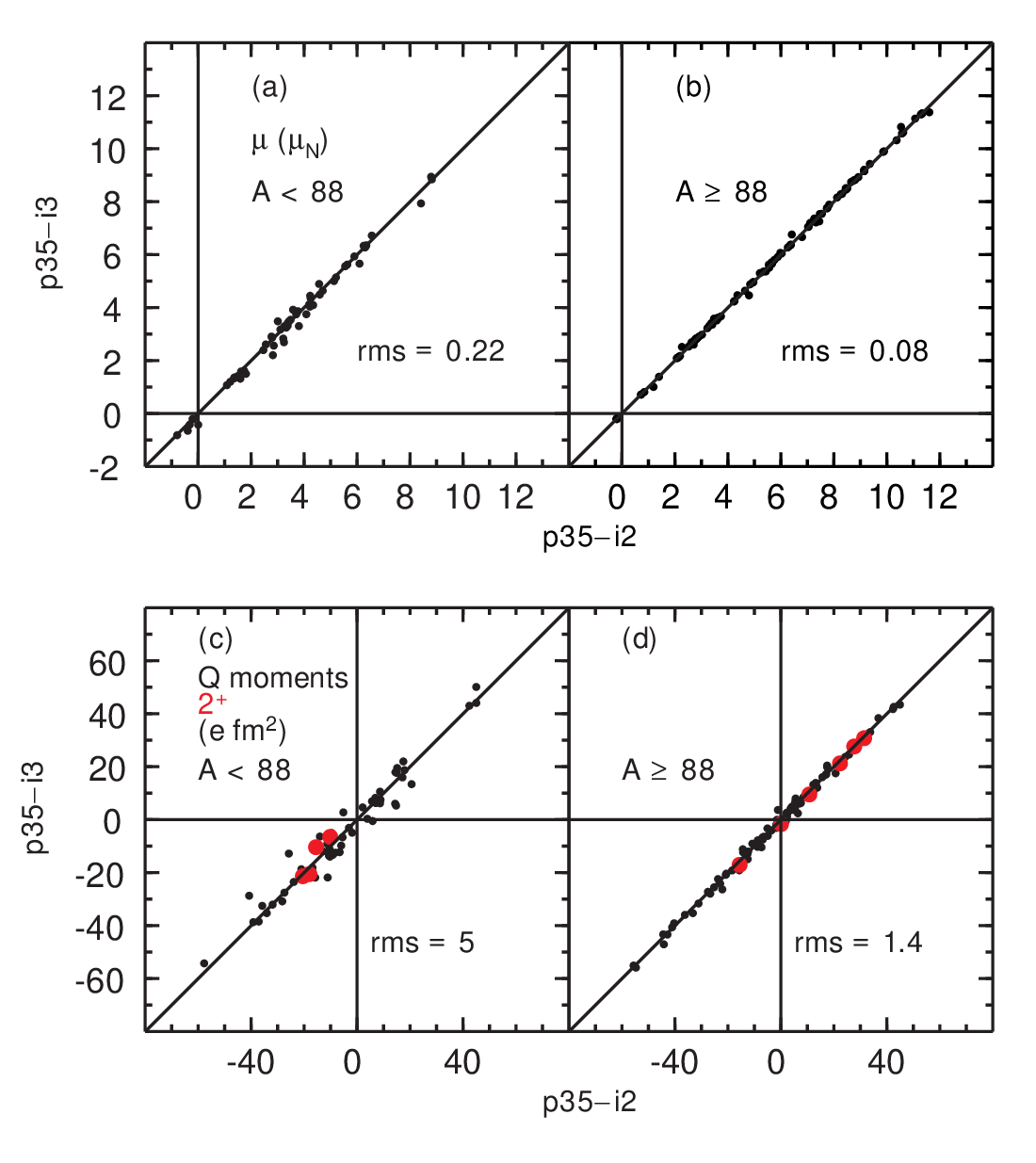}
\caption{Comparison of  moments obtained with the
p35-i2 and p35-i3 Hamiltonians.
}
\label{(3)}
\end{figure}

\begin{figure}
\includegraphics[scale=0.40]{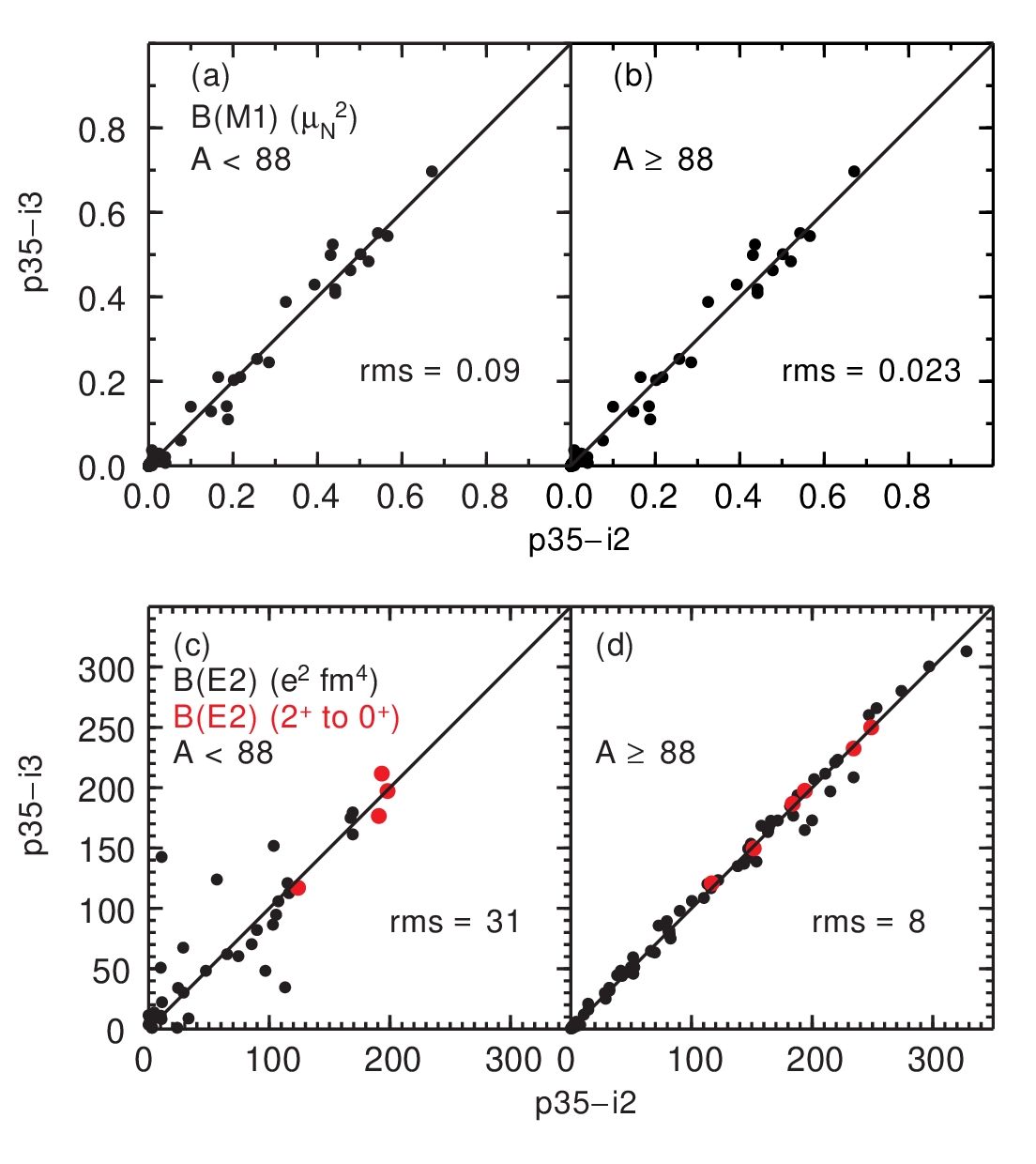}
\caption{Comparison of $B(M1)$ and $B(E2)$ values obtained with the
p35-i2 and p35-i3 Hamiltonians.
}
\label{(4)}
\end{figure}

We show the comparison of results obtained with the IMSRG(2) and IMSRG(3f2) starting Hamiltonians for 35 VLC, (p35-i2) and (p35-i3), respectively, in Figs. \ref{(3)} and \ref{(4)}. The scatter and RMS deviations are similar to the (p30-i2) and (p35-i3) comparisons discussed above.

\begin{figure}
\includegraphics[scale=0.40]{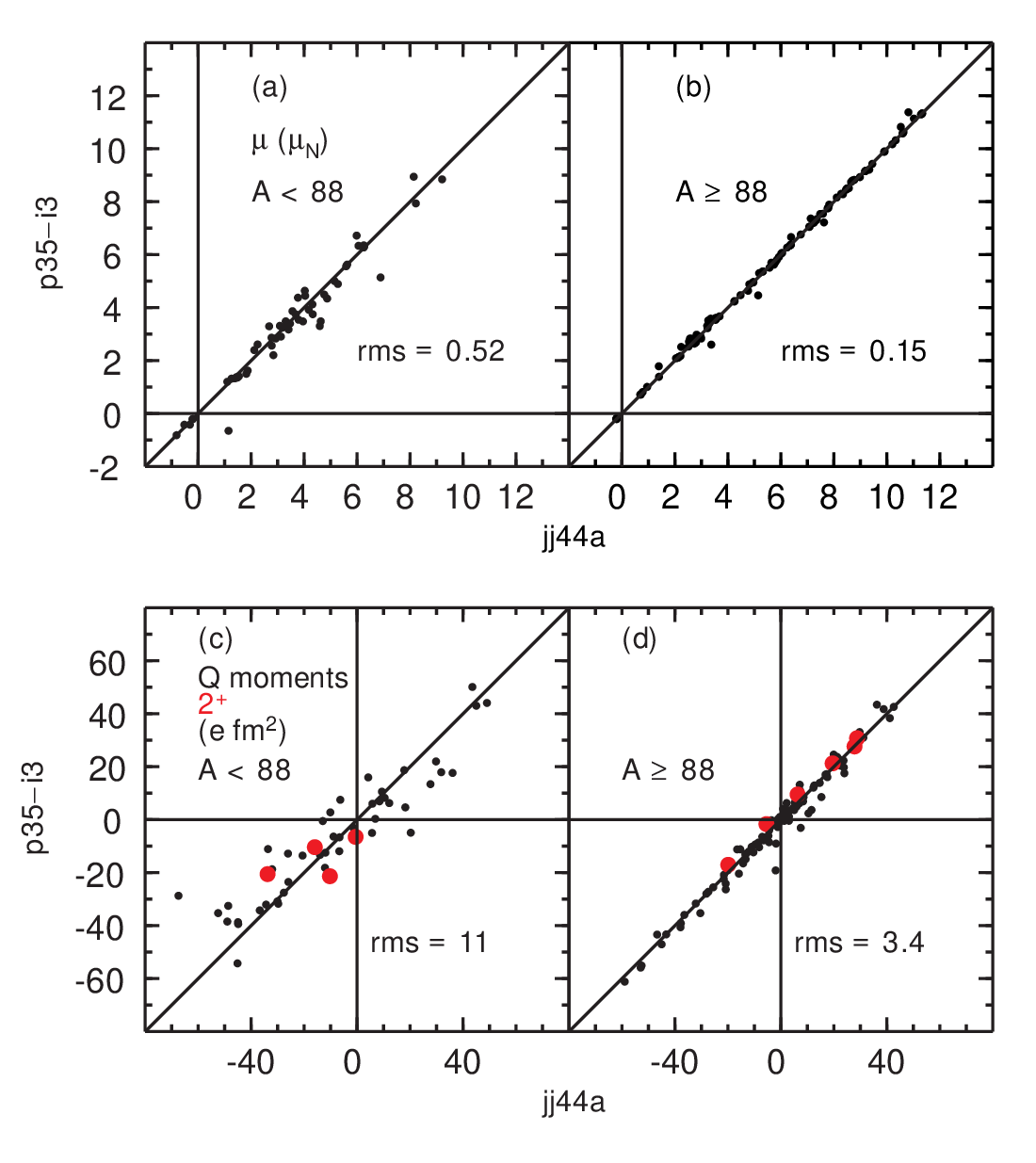}
\caption{Comparison of  moments obtained with the
jj44a and p35-i3 Hamiltonians.
}
\label{(5)}
\end{figure}

\begin{figure}
\includegraphics[scale=0.40]{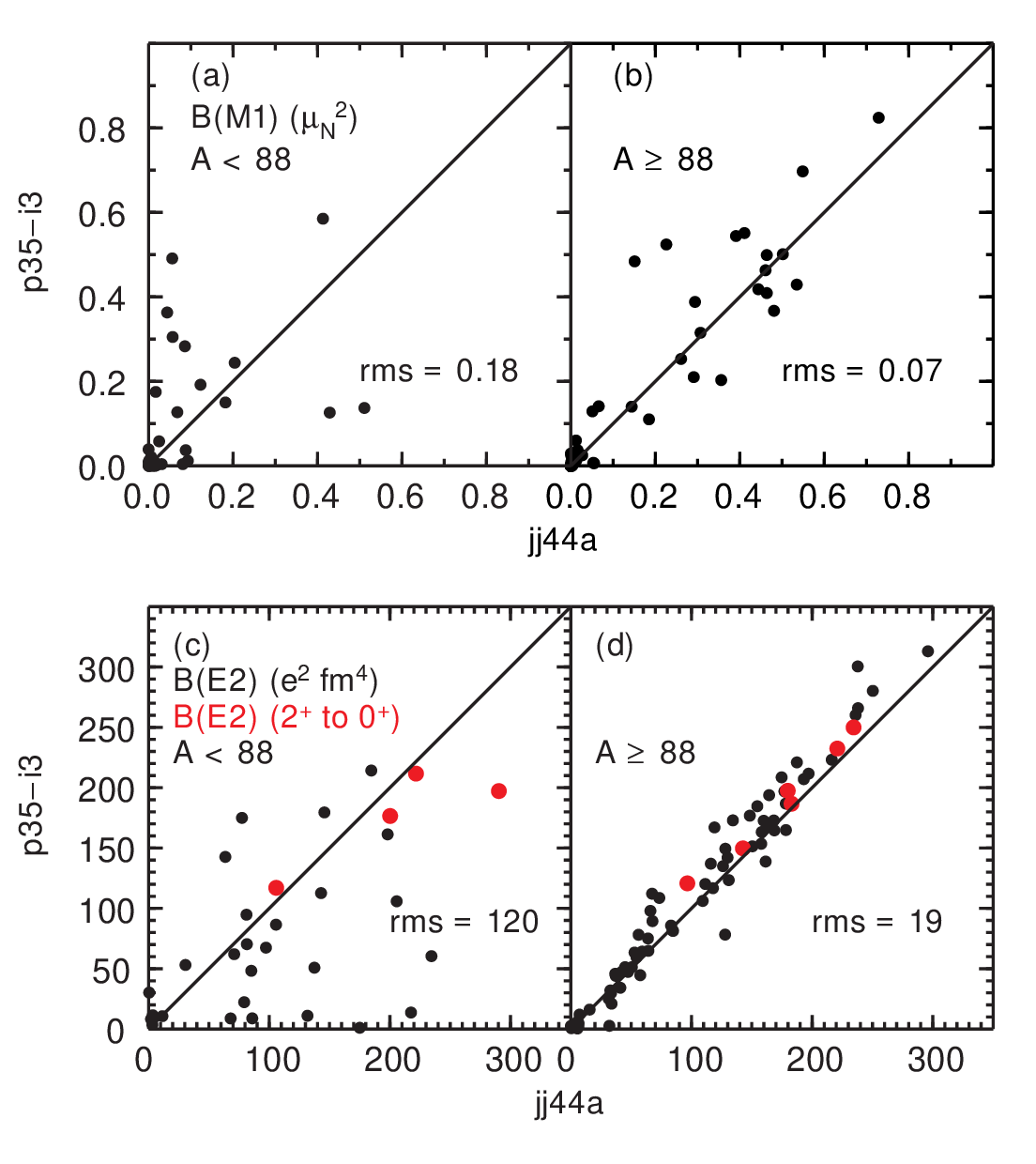}
\caption{Comparison of $B(M1)$ and $B(E2)$ values obtained with the
jj44a and p35-i3 Hamiltonians.
}
\label{(6)}
\end{figure}

The jj44a Hamiltonian \cite{jj44a} was obtained from a fit to the $  N=50 $ energy data known in 2004. 
Comparisons of electromagnetic results obtained with jj44a and p35-i3 are shown in Figs. \ref{(5)} and \ref{(6)}. 
The scatter is significantly larger compared to those in Figs. (1-4), 
particularly for the $B(M1)$ and $B(E2)$ values shown in Fig. \ref{(6)}.
The comparisons between jj44a and p35-i3 are good examples of
the sensitivity of electromagnetic data to the Hamiltonian.

\section{Comparison between theory and experiment}
Results for the p35-i2 and p35-i3 Hamiltonians are very similar, and we
will only give results from p35-i3.
We compare experimental data with those obtained from p35-i3 and from the older Hamiltonians 
n50j \cite{jw88} and jj44a \cite{jj44a}. 

\begin{figure}
\includegraphics[scale=0.5]{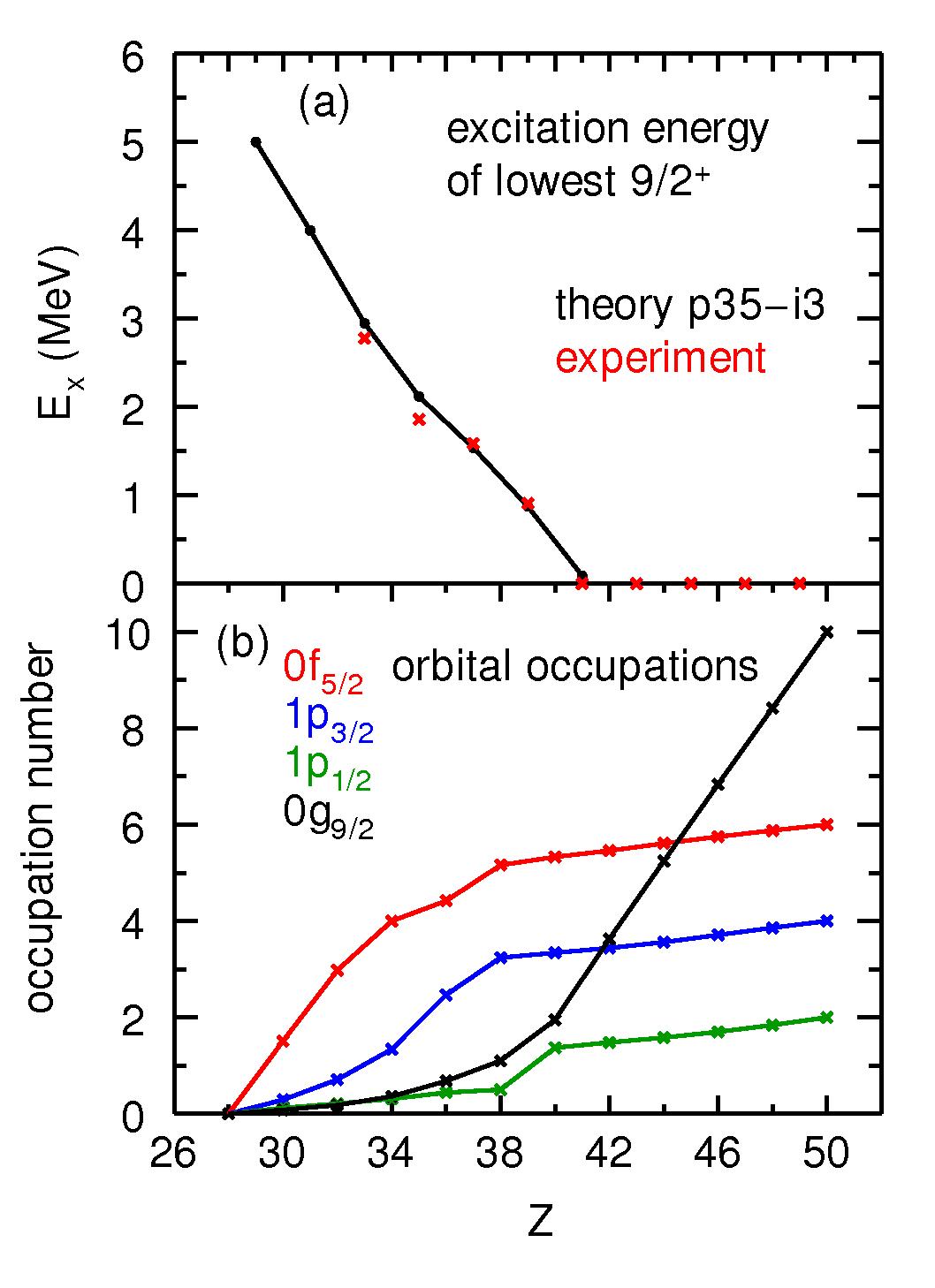}
\caption{The top panel (a) shows the excitation energy of the lowest 9/2$^+$ states
for odd-even nuclei obtained with the p35-i3 Hamiltonian compared
to experiment.
The bottom panel (b) shows the orbital occupations numbers 
for the ground states of even-even nuclei obtained with the p35-i3 Hamiltonian.
}
\label{occ}
\end{figure}

The OBTD in Eq. (2) are 
determined by the orbital occupancies in the initial 
and final wavefunctions. The orbital occupations numbers
for the ground states of even-even nuclei obtained with the p35-i3 Hamiltonian
are shown in the bottom panel of Fig.  \ref{occ}.
Spectroscopic factors extracted from $(e,e'p)$ reactions
provide an experimental test of these orbital occupations.
The experimental Spectroscopic factors extracted for $^{90}Zr(e,e'p)^{89}$Y \cite{zr90eep}
are 0.72(7) for 1/2$^{-}$, 0.43(5) for 9/2$^{+}$, 1.86 for 3/2$^{-}$, and 2.77(19) for 5/2$^{-}$,
to be compared with the p35-i3 values of 0.71, 0.77, 1.94 and 2.95,
respectively. The theoretical spectroscopic factors have been
multiplied by the typical quenching factor of 0.60 observed for
nuclei near stability \cite{gade2014}. Many of the electromagnetic 
observables discussed below reflect the sequential filling of
the orbitals observed in Fig.\ref{occ}. The top panel of Fig. \ref{occ}
shows how the energy of the 9/2$^{+}$ level associated with the $0g_{9/2}$ orbital 
comes down in energy until it becomes the ground state at $Z=41$.
The systematic filling of the $0g_{9/2}$ orbital above $Z=40$ gives
rise to the $j=9/2$ seniority patterns in the spectra and decays discussed below.

\subsection{Magnetic Moments}
\begin{figure}
\includegraphics[scale=0.35]{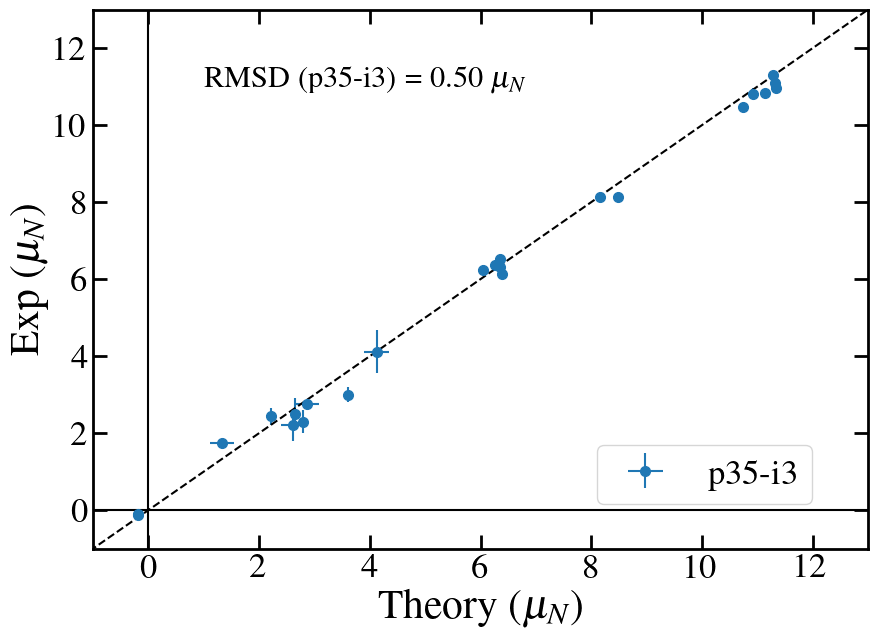}
\caption{Comparison of magnetic dipole moments values between experiment and theory for the p35-i3 Hamiltonians
}
\label{dipoles}
\end{figure}

Experimental magnetic dipole moments are compared to the calculations in Table \ref{DipoleTable},
Fig. \ref{dipoles}, and the upper panels of Figs. \ref{muQgs} and \ref{muQexcited}.
The agreement between experiment and theory is good with an rms deviation
of 0.50 $\mu_N)$ which is a factor of two larger than the theoretical Hamiltonian uncertainty of 0.24 from
Fig. 1 and 3. The experimental moment for $^{81}$Ga is larger that all 
of the calculations. This may reflect an additional orbital dependence 
of the effective $g$ factors for the $0f_{5/2}$ orbital. More experimental
data is needed below $A \leq 85$.

The low-lying magnetic moments for odd-even nuclei in the $\pi j4$ model space are shown in the upper panels of Fig. \ref{muQgs}, 
and for even-even nuclei in the upper panel of Fig. \ref{muQexcited}; 
the trends of these figures are discussed in Sec. IV.D. 

being the subject of a subsequent discussion on $j^n$ 
configurations following electromagnetic transitions.

\begin{table}[]
\centering
\caption[Experimental and theoretical dipole values]{Magnetic dipole moments ($\mu_N)$} 
\renewcommand{\arraystretch}{1.4}
\begin{tabular}{cccccc}
\toprule\toprule
Nucleus  & $J^\pi$ & Exp & p35-i3 & jj44a & n50j \\
\midrule
$^{79}$Cu & $5/2^-$ &   &   1.329  & 1.329  &   1.329 \\

$^{80}$Zn &  $2^+$  &   &   1.317  & 1.256  &   1.236 \\

$^{81}$Ga & $5/2^-$ & 1.747(5)\textsuperscript{\cite{quadGa81}}  &   1.328  & 1.382  &   1.330   \\

$^{82}$Ge &  $2^+$  &   &   1.501  & 1.818  &   1.304 \\

$^{83}$As & $5/2^-$ &   &   1.361  & 1.487 &   1.313 \\

$^{84}$Se &  $2^+$  &   &   2.395  & 2.118  &   1.792   \\

$^{85}$Br & $3/2^-$ &   &   3.295  & 2.681 &   3.136\\

$^{86}$Kr &  $2^+$  & 2.2(4)\textsuperscript{\cite{mu86Kr}}  &   2.610  & 2.243  &   2.175    \\

$^{87}$Rb & $3/2^-$ &  2.75131(12)\textsuperscript{\cite{mu87Rb}} & 2.86469  &   2.77036   &  3.10328   \\

$^{88}$Sr & $2^+$ &  2.44(22)\textsuperscript{\cite{mu88Sr}}    & 2.17  &   2.21   &   2.43 \\

$^{89}$Y & $1/2^-$ &  -0.1374154(3)\textsuperscript{\cite{mu89Y1}}    & -0.17814 &   -0.20371 & -0.17663 \\

$^{89}$Y & $9/2^+$ &  6.37(4)\textsuperscript{\cite{mu89Y9}}    & 6.27 &  6.23   &   6.19 \\

$^{90}$Zr &  $2^+$ &  2.5(4)\textsuperscript{\cite{mu90Zr2-3}}     & 2.6 &  2.7  &   2.6 \\ 

$^{90}$Zr &  $5^-$ &  6.25(13)\textsuperscript{\cite{mu90Zr5}}     & 6.04 &  6.02   &   5.59\\ 

$^{90}$Zr &  $3^-$ &  3.0(2)\textsuperscript{\cite{mu90Zr2-3}}    & 3.6 &  3.3   &   3.4 \\ 

$^{90}$Zr &  $8^+$ &  10.84(6)\textsuperscript{\cite{mu90Zr8}}     & 11.14  &  11.02   &   10.85 \\ 

$^{91}$Nb &  $9/2^+$ &  6.521(2)\textsuperscript{\cite{mu91Nb9-1}}    & 6.346 &  6.346   &   6.346 \\

$^{91}$Nb &  $1/2^-$ &  -0.101(2)\textsuperscript{\cite{mu91Nb9-1}}    & -0.192  &  -0.205   &   -0.201  \\

$^{91}$Nb &  $(13/2)^-$ &  8.14(13)\textsuperscript{\cite{mu91Nb13}}    & 8.16  &  8.11  &   7.64 \\

$^{91}$Nb &  $(17/2)^-$ &  10.82(14)\textsuperscript{\cite{muN50iso}}    & 10.92  &  10.89  &   10.42 \\

$^{92}$Mo & $2^+$ &  2.3(3)\textsuperscript{\cite{mu92Mo2}}     & 2.8  &  2.8   &   2.7   \\

$^{92}$Mo & $8^+$ &  11.30(5)\textsuperscript{\cite{muN50iso}}     & 11.29  &  11.29   &   11.29   \\

$^{93}$Tc &  $9/2^+$ &  6.32(6)\textsuperscript{\cite{mu93Tc9}}   & 6.36  &  6.36   &   6.36 \\

$^{93}$Tc &  $(17/2)^-$ &  10.46(5)\textsuperscript{\cite{muN50iso}}   & 10.75  &  10.46   &   10.46 \\

$^{94}$Ru  & $2^+$  &   &   2.790 &  2.812 &   2.802\\

$^{94}$Ru  & $6^+$  & 8.124(48)\textsuperscript{\cite{muN50iso}}  &   8.477 &  8.477 &   8.487\\

$^{94}$Ru  & $8^+$  & 11.096(40)\textsuperscript{\cite{muN50iso}}  &   11.311 &  11.311 &   11.321\\

$^{95}$Rh  &  $9/2^+$   &   &   6.369 &   6.364 &  6.369 \\

$^{96}$Pd  &  $(8^+)$   &   10.97(6)\textsuperscript{\cite{mu96Pd8}}    &   11.33  &   11.33    &  11.33 \\

$^{97}$Ag  &  $9/2^+$   & 6.13(12)\textsuperscript{\cite{mu97Ag}}  &   6.375  &  6.375     &  6.375 \\

$^{98}$Cd  &  $2^+$ &    &  2.799 &   2.823    &  2.847 \\

$^{99}$In  & $9/2^+$    &   &   6.375 &   6.375 &  6.375   \\

\bottomrule\bottomrule
\label{DipoleTable}
\end{tabular}
\end{table}


\subsection{Quadrupole Moments}

\begin{figure}
\includegraphics[scale=0.35]{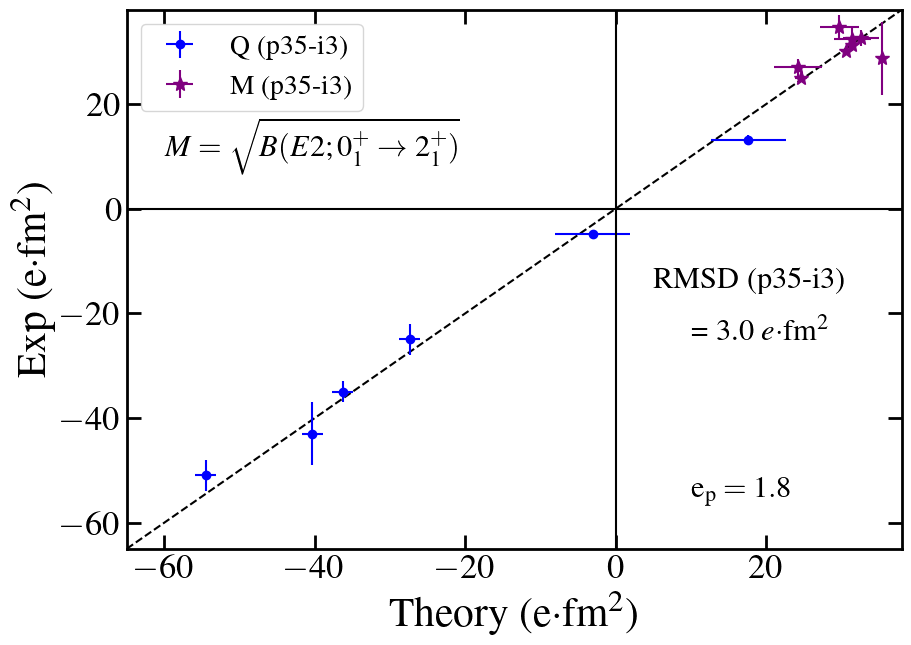}
\caption{Comparison of quadrupole moment Q and $M = \sqrt{B(E2)}$ between experiment and theory using an effective charge e$_\text{p} = 1.8$. For the $M$ values plotted, we input $B(E2; 0^+_1 \rightarrow 2^+_1)$.
}
\label{qm}
\end{figure}

\begin{figure}
\includegraphics[scale=0.35]{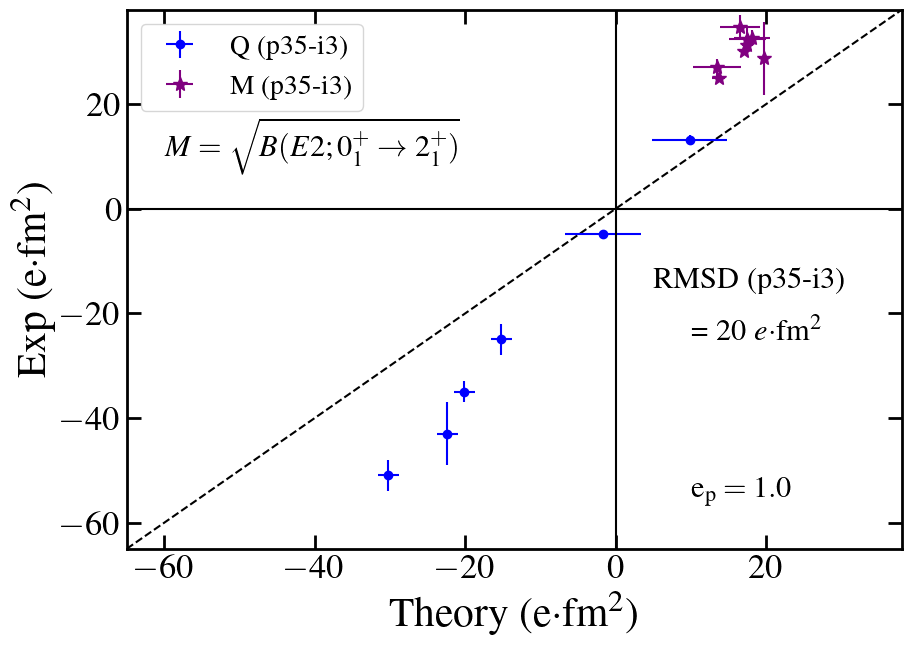}
\caption{Comparison of quadrupole moment Q and $M = \sqrt{B(E2)}$ between experiment and theory for the "free charge" model (e$_p = 1.0$). For the $M$ values plotted, we input $B(E2; 0^+_1 \rightarrow 2^+_1)$
}
\label{qmfree}
\end{figure}

\begin{figure}
\includegraphics[scale=0.22]{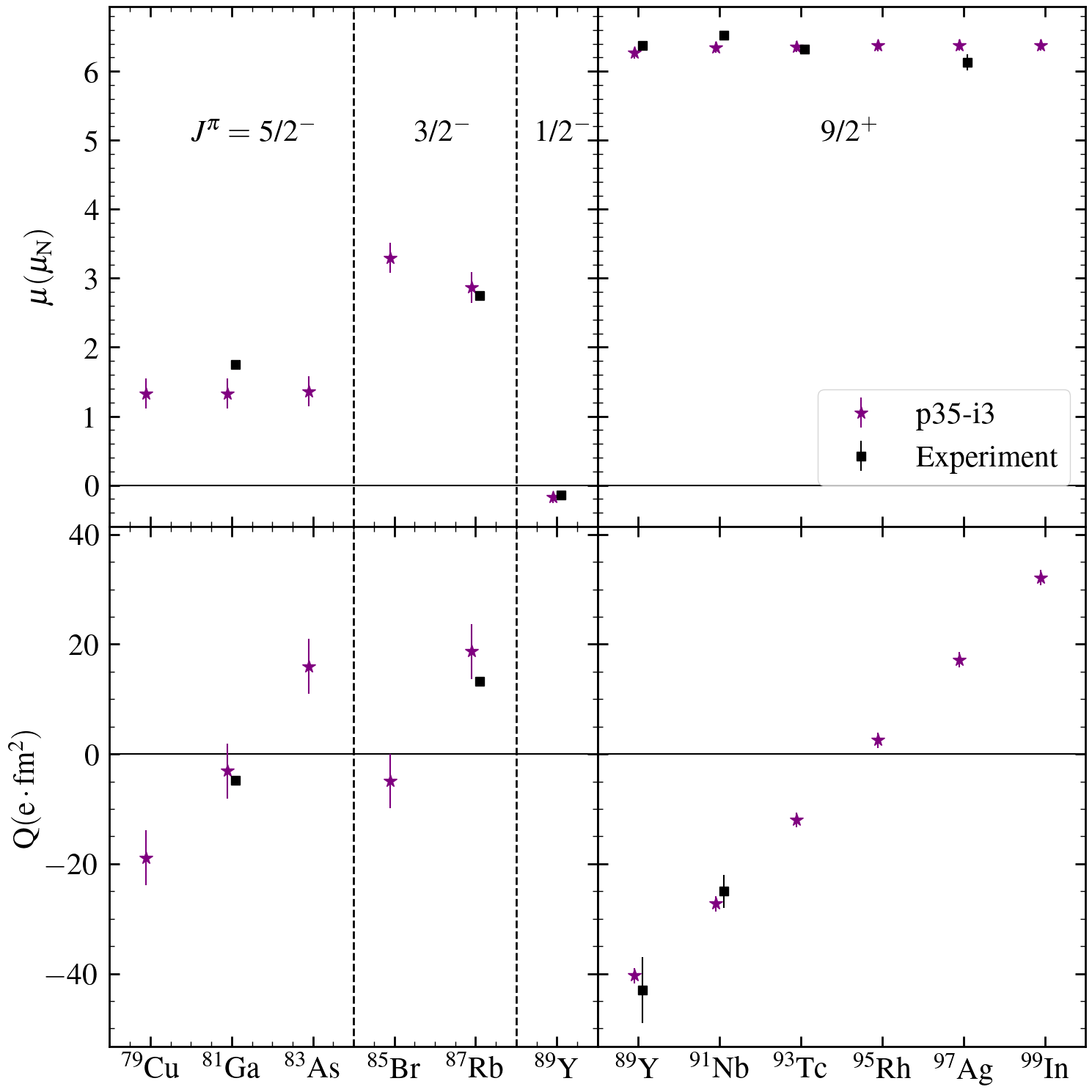}
\caption{The upper two panels show the magnetic dipole moments and bottom two panels the electric quadrupole moments
of the lowest lying instance of each spin for the odd-$Z$ nuclei. The shown values correspond to the ground states of most nuclei, though the $9/2^+$ in $^{89}$Y is the first excited state. Values of $Q$ were calculated using an effective charge e$_\text{p} = 1.8$.
}
\label{muQgs}
\end{figure}

Experimental electric quadrupole moments are compared to the calculations in Table \ref{QuadrupoleTable},
Fig. \ref{qm}, and the lower panels of Figs. \ref{muQgs} and \ref{muQexcited}.
In Fig. \ref{qm} the quadrupole moments are compared to the matrix $M = \sqrt{B(E2)}$ for $0^+_1 \rightarrow 2^+_1$ transitions.
This figure shows that the effective charge of $e_p=1.8$ is appropriate for both
quadrupole moments and electromagnetic transitions. 
The rms deviation of 3.0 e-fm between experiment and theory in Fig. \ref{qm} is
about the same as the theoretical uncertainty of 5 e-fm shown in Figs. 1 and 3.
Fig. \ref{qmfree} shows the comparison with experiment using
the free-nucleon value of $e_p=1$. The theory is rotated downward by a factor 1.8 compared to
experiment with a much larger rms deviation of 20 e-fm between experiment and theory.

The lower panels of Fig. \ref{muQgs} show the mass dependence of
the quadrupole moments for the  odd-even nuclei in the $\pi j4$ model space. 
As one scans this figure from left to right, 
it is divided into regions where the low-lying states are dominated by a given $j^\pi$ orbital, 
where one can see the transition from oblate to prolate as one fills the shell with protons. 
This is explained by an increase in the orbital occupation numbers for the low-lying states of 
these nuclei as we add protons. 

\begin{table}[]
\centering
\caption[Experimental and theoretical quadrupole moment values $(Q)$]{Experimental quadrupole moment values (e$\cdot$fm$^2$) and the corresponding calculations 
from the new p35-i3 and older jj44a and n50j Hamiltonians for A = 79 - 99.}
\renewcommand{\arraystretch}{1.4}
\begin{tabular}{cccccc}
\toprule\toprule
Nucleus  & $J^\pi$ & Exp (e$\cdot$fm$^2$) & p35-i3 & jj44a & n50j \\
\midrule
$^{79}$Cu & $5/2^-$ &   &   -18.91  & -18.91  &   -18.91 \\

$^{80}$Zn &  $2^+$  &   &   -6.50 & -0.56  &   0.77 \\

$^{81}$Ga & $5/2^-$ & -4.8(8)\textsuperscript{\cite{quadGa81}}  &   -3.09  & -1.40  &   -1.27  \\

$^{82}$Ge &  $2^+$  &   &   -21.36  & -10.27  &   -18.62 \\

$^{83}$As & $5/2^-$ &   &   15.95  & 4.20 &   16.95\\

$^{84}$Se &  $2^+$  &   &   -20.56  & -33.80  &   -19.08   \\

$^{85}$Br & $3/2^-$ &   &   -4.93  & 20.26 &   -8.23 \\

$^{86}$Kr &  $2^+$  &   &   -10.39  & -15.95  &   -21.50    \\

$^{87}$Rb & $3/2^-$ &  13.2(1)\textsuperscript{\cite{quad87Rb}} & 18.68  &   17.80   &  14.15   \\

$^{88}$Sr & $2^+$ &    & 27.63 &   27.79   &   21.53 \\

$^{89}$Y & $9/2^+$ &  -43(6)\textsuperscript{\cite{mu89Y1}}    & -40.62 &   -37.91 & -33.67 \\

$^{90}$Zr &  $8^+$ &  -51(3)\textsuperscript{\cite{mu90Zr8}}     & -55.09  &  -52.62   &   -51.59 \\ 

$^{91}$Nb &  $9/2^+$ &  -25(3)\textsuperscript{\cite{mu91Nb9-1}}    & -27.24 &  -27.40   &   -27.24 \\

$^{92}$Mo & $8^+$ &  -34(2)\textsuperscript{\cite{quad92Mo}}     & -35.96  &  -36.46   &   -32.53   \\

$^{93}$Tc &  $9/2^+$ &   & -12.00  &  -12.10   &   -9.31 \\

$^{94}$Ru  & $8^+$  &   &   -9.39  &  -9.73 &   -7.27 \\

$^{95}$Rh  &  $9/2^+$   &   &   2.55 &   2.65 &  4.07 \\

$^{96}$Pd  &  $8^+$   &   &   17.00 &   16.83   &  17.42 \\

$^{97}$Ag  &  $9/2^+$   &   &   17.18  &  17.23     &  17.68 \\

$^{98}$Cd  &  $8^+$ &    &  42.59 &   42.59   &  42.59 \\

$^{99}$In  & $9/2^+$    &   &   32.12 &   32.12 &  32.12   \\

\bottomrule\bottomrule
\label{QuadrupoleTable}
\end{tabular}
\end{table}


\subsection{Electromagnetic Transitions}
\begin{figure}
\includegraphics[scale=0.37]{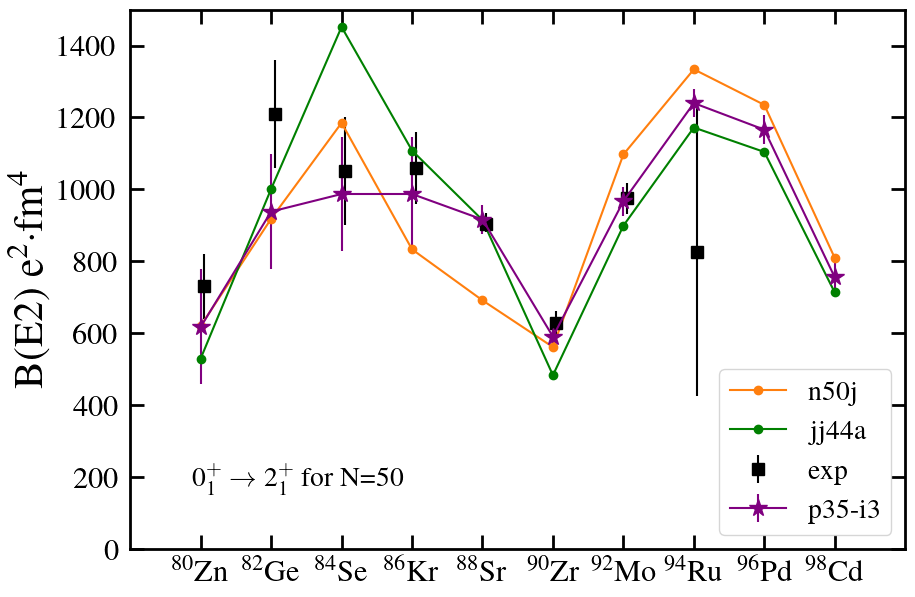}
\caption{The $B(E2)$ values for the $0^+_1 \rightarrow 2^+_1$ transition in the even-even 
nuclei within the $\pi j4$ model space. 
The theoretical values were calculated using an effective charge of $e_p = 1.8$ for the $E2$ operator.
}
\label{be02}
\end{figure}

Shown in Fig. \ref{be02} are the $B(E2; 0^+_1 \rightarrow 2^+_1)$ for 
all of the even-even nuclei in the $\pi j4$ model space, 
while Fig. \ref{be86} shows $B(E2; 8^+_1 \rightarrow 6^+_1)$ for even-even nuclei with $Z\geq 40$.  
As mentioned in the previous section, we adopt the effective charge of $e_\text{p} = 1.8$ for our $E2$ operator. 
One can see that the p35-type Hamiltonian generally outperforms the n50j and jj44a Hamiltonians 
for the $0^+_1 \rightarrow 2^+_1$ transitions and performs 
comparably well to the n50j Hamiltonian for the $8^+_1\rightarrow 6^+_1$ transitions 
when compared to the experimental data in this region of nuclei. 
There is a large scatter between the various model predictions for the 
$8^+_1 \rightarrow 6^+_1$ transition in $^{90}$Zr, and jj44a agrees closely with the experimental results. 
Table \ref{BE2table} details a larger comparison of transitions between theory and experiment 
and includes the results of an additional seniority model in the final columns \cite{seniority}.

\begin{table*}
\caption[Experimental and theoretical $M1$ transition strengths $B(M1)$]
{Experimental values for $M1$ transition strengths and the corresponding calculations from the new p35-i3 and older jj44a and n50j Hamiltonians.
Experimental data are all from  the ENSDF database as of January 1st, 2024,
http://www.nndc.bnl.gov/ensarchivals/.
}\centering 
\renewcommand{\arraystretch}{1.4}
\begin{tabular}{ccccccc}
\toprule\toprule
&&&\multicolumn{3}{c}{$\hspace{2.5cm}B(M1; J^\pi_i \rightarrow{} J^\pi_f) (\mu_N^2) $} \\
\cmidrule{4-7}
Nucleus  & $J^\pi_i$  & $J^\pi_f$ & Exp & p35-i3 & jj44a & n50j  \\
\midrule
$^{86}$Kr &   2$^+_2$      &  2$^+_1$    &   0.154(20)       & 0.165  & 0.131 & 0.39\\

$^{87}$Rb &   5/2$^-_1$      &  3/2$^-_1$    &   0.008(5)    & 0.0094 & 0.014 & 742\\

$^{87}$Rb &   1/2$^-_1$      &  3/2$^-_1$    &   0.64(7)     & 1.04  & 0.42 & 960\\

$^{88}$Sr &   2$^+_2$      &  2$^+_1$    &   0.074(5)        & 0.078  & 0.033 & 0.014\\

$^{88}$Sr &   1$^+_1$      &  0$^+_1$    &   0.33(3)         & 0.43 & 0.54 & 0.26\\

$^{88}$Sr &   3$^+_1$      &  2$^+_1$    &   0.008(4)        & 0.010 & 0.007 & 0.00014\\

$^{88}$Sr &   3$^+_1$      &  2$^+_2$    &   0.027(5)       & 0.046  &  0.027 & 0.0023 \\

$^{88}$Sr  &   2$^+_3$       &  2$^+_2$    &   0.09(2)       & 0.041  & 0.021 &  0.91 \\

$^{88}$Sr  &   3$^-_2$       &  3$^-_1$    &   0.014(4)       & 0.016  & 0.006 &  0.051\\
\bottomrule\bottomrule
\end{tabular}
\label{BM1table}
\end{table*}

\begin{table*}
\caption[Experimental and theoretical E2 transition strengths]{Experimental values for 
$E2$ transitions and the corresponding calculations from the 
new p35-i3 and older jj44a and n50j Hamiltonians with $e_p = 1.8$e, 
as well as those from a seniority model \cite{seniority} (denoted $\hat{T}_1$). 
States with seniority $\upsilon$ correspond to $\hat{T}_1$ calculations.}
\centering 
\renewcommand{\arraystretch}{1.4}
\begin{tabular}{cccccccccc}
\toprule\toprule
&&&\multicolumn{3}{c}{$\hspace{2.5cm}B(E2; J^\pi_i \rightarrow{} J^\pi_f) (e^2$ fm$^4)$} \\
\cmidrule{4-8}
Nucleus  & $J^\pi_i$  & $J^\pi_f$ & Exp & p35-i3 & jj44a & n50j & $\hat{T}_1(E2)$ & $\upsilon_i$ & $\upsilon_f$ \\
\midrule
$^{80}$Zn &   0$^+_1$      &  2$^+_1$    &   730(90)\textsuperscript{\cite{be80Zn}}            & 585  & 529 & 518\\

$^{82}$Ge &   0$^+_1$      &  2$^+_1$    &   1210(150)\textsuperscript{\cite{be2Gade}}            & 883 & 1001 & 742\\

$^{84}$Se &   0$^+_1$      &  2$^+_1$    &   1050(150)\textsuperscript{\cite{be2Gade}}            & 986  & 1452 & 960\\

$^{86}$Kr &   0$^+_1$      &  2$^+_1$    &   1060(100)\textsuperscript{\cite{be2Kr86}}            & 1058  & 1108 & 675\\

$^{88}$Sr &   0$^+_1$      &  2$^+_1$    &   903$^{(+32)}_{(-23)}$\textsuperscript{\cite{be2PBS}}    & 933 & 915 & 560\\

$^{90}$Zr &   0$^+_1$      &  2$^+_1$    &   627(34)\textsuperscript{\cite{be2PBS}}            & 604 & 483 & 454\\

$^{92}$Mo &   0$^+_1$      &  2$^+_1$    &   975(43)\textsuperscript{\cite{be2PBS}}           & 987  &  899 & 889 & 445  &   0  &  2 \\

$^{94}$Ru  &   0$^+_1$       &  2$^+_1$    &   825(400)\textsuperscript{\cite{be2ru94}}           & 1250  & 1172 &  919 & 680 &   0&  2\\

$^{96}$Pd  &   0$^+_1$   &  2$^+_1$       &                     & 1162   & 1104  & 1000\\

$^{98}$Cd  &   0$^+_1$   &  2$^+_1$       &                     & 749   & 713  & 655\\

$^{92}$Mo  &   4$^+_1$       &  2$^+_1$    &   132$^{(+7)}_{(-6)}$\textsuperscript{\cite{seniority}}   & 135  &  126 & 87 & 103 &   2&  2\\

$^{92}$Mo  &   6$^+_1$       &  4$^+_1$    &   81(2)\textsuperscript{\cite{seniority}}            & 81 &  85 & 72 & 71  &   2&  2\\

$^{92}$Mo  &   8$^+_1$      &  6$^+_1$    &   28.6(3)\textsuperscript{\cite{seniority}}           & 32.0 &  32.7 & 28.9 & 28  &   2&  2 \\

$^{93}$Tc  &   17/2$^+_1$    &  13/2$^+_1$ &   88(18)\textsuperscript{\cite{seniority}}            & 117  &  117 & 96 & 99  &   3&  3\\

$^{93}$Tc  &   21/2$^+_1$   &  17/2$^+_1$ &   73(5)\textsuperscript{\cite{seniority}}             & 61   & 66 &  60 & 57  &   3&  3 \\

$^{94}$Ru  &   4$^+_1$       &  2$^+_1$    &   38(3), 103(24)\textsuperscript{\cite{seniority}}    & 12.0  & 7.2 & 1.7 &  12  &   2&  2\\

$^{94}$Ru  &   6$^+_1$      &  4$^+_1$    &   3.0(2)\textsuperscript{\cite{seniority}}            & 5.0  & 6.3 &  4.8 & 8.0 &   2 &  2\\

$^{94}$Ru  &   8$^+_1$      &  6$^+_1$    &   0.09(1)\textsuperscript{\cite{seniority}}           & 1.42 &  1.98 & 0.73 & 3.2 &   2&  2 \\

$^{95}$Rh  &   21/2$^+_1$   &  17/2$^+_1$ &   24(2)\textsuperscript{\cite{seniority}}             & 4.992 &  0.004 & 0.23 & 0   &   3&  3 \\

$^{95}$Rh  &   21/2$^+_1$   &  17/2$^+_2$ &   113(13)\textsuperscript{\cite{seniority}}           & 185   & 200 &  194 & 138 &   3&  5 \\

$^{96}$Pd  &   $\left(6^+_1 \right)$   &  $\left (4^+_1\right)$ &   20.4(2.9)\textsuperscript{\cite{be2Pd96}}  & 16.1   & 15.5 & 16.8 \\

$^{96}$Pd  &   $\left(8^+_1 \right)$   &  $\left (6^+_1\right)$ &   9.4(1.0)\textsuperscript{\cite{be2Pd96}}  & 6.0   & 5.4 & 7.4 \\

$^{98}$Cd  &   $\left(4^+_1 \right)$   &  $\left (2^+_1\right)$ &   98(50)\textsuperscript{\cite{be2Cd98}}  & 165   & 169 & 169 \\

$^{98}$Cd  &   $\left(6^+_1 \right)$   &  $\left (4^+_1\right)$ &   110(5)\textsuperscript{\cite{be2Cd98}}  & 117   & 118 & 118 \\

$^{98}$Cd  &   $\left(8^+_1 \right)$   &  $\left (6^+_1\right)$ &   39(4)\textsuperscript{\cite{be2Cd98}}  & 47   & 47 & 47 \\

\bottomrule\bottomrule
\end{tabular}
\label{BE2table}
\end{table*}


\subsection{Configuration with $j^n$}
\begin{figure}
\includegraphics[scale=0.30]{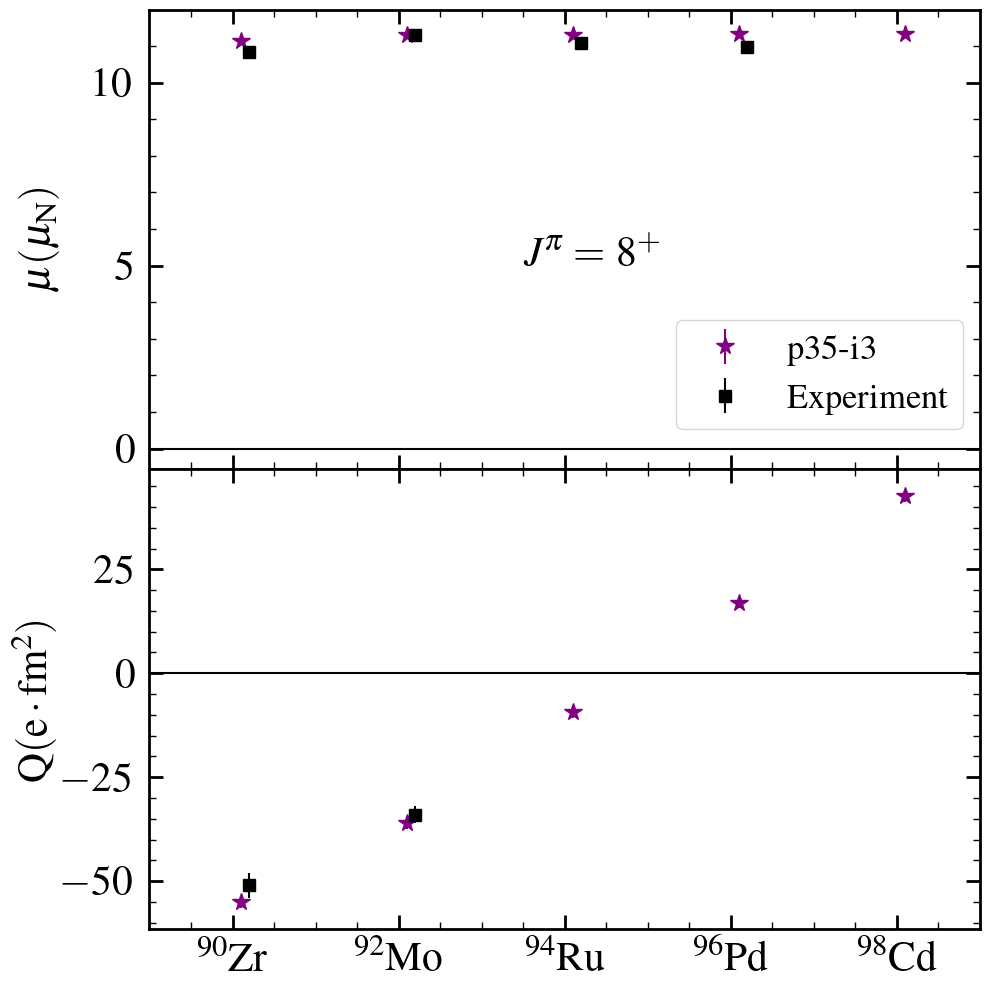}
\caption{The upper panel shows the magnetic dipole moments and bottom panel shows the 
electric quadrupole moments of the $8^+_1$ isomeric states for the even-even nuclei. Values of $Q$ were 
calculated using an effective charge $e_p = 1.8$.
}
\label{muQexcited}
\end{figure}

An interesting observation in Figs. \ref{muQgs} and \ref{muQexcited} is that one 
can see that for a given $J^\pi$ value, the magnetic moments are uniform. 
This uniformity is due to the relative purity of the wavefunctions: 
All valence particles are in relatively pure configurations for these states, 
with configurations of the form $[nlj]^n$, and with relatively little mixing of 
particle-hole configurations. 
The purity of these configurations combined with the fact that the strong interaction 
pairs spins to zero means that the total spin of the nucleus for each odd-even nucleus 
will always be the spin of the single unpaired proton in a given single-particle orbital.

Looking at the results of several Hamiltonians in Fig. \ref{muQgs}, one can see that the 
low-lying states in the nuclei with $A \leq 83$ orbits are well described by pure 
configurations of $\left[0f_{5/2}\right]^n$, while the low-lying states in nuclei with $A \geq 91$ 
are well described by pure configurations of $\left[0g_{9/2} \right]^n$. The increased scatter 
in the dipole moments for $^{85}$Br and $^{87}$Rb is due to an increased configuration mixing 
in their wavefunctions, resulting in reduced purity of the configurations contained within.
The $J^\pi = 1/2^-$ ground state of $^{89}$Y involves a more substantial configuration mixing, 
in particular, with the $0g_{9/2}$ orbit having an expectation value of 1 proton occupying its orbit.

If we consider the $8^+$ states in Fig. \ref{muQexcited}, we see a level of uniformity similar 
to that of the low-lying states in odd-even nuclei with $A \leq 83$ and $A \geq 91$, 
this time spanning the even-even nuclei with $A \geq 90$. The underlying reason for this 
uniformity is identical to the reason for uniformity in the odd-even nuclei: 
These wavefunctions involve relatively pure configurations of the form $\left[0g_{9/2}\right]^n$.
What makes these states different from the low-lying ones in the odd-even nuclei is that these 
states are excitations that involve pair breaking in the $0g_{9/2}$ orbit, making these seniority $\nu=2$ excitations.

\begin{figure}
\includegraphics[scale=0.25]{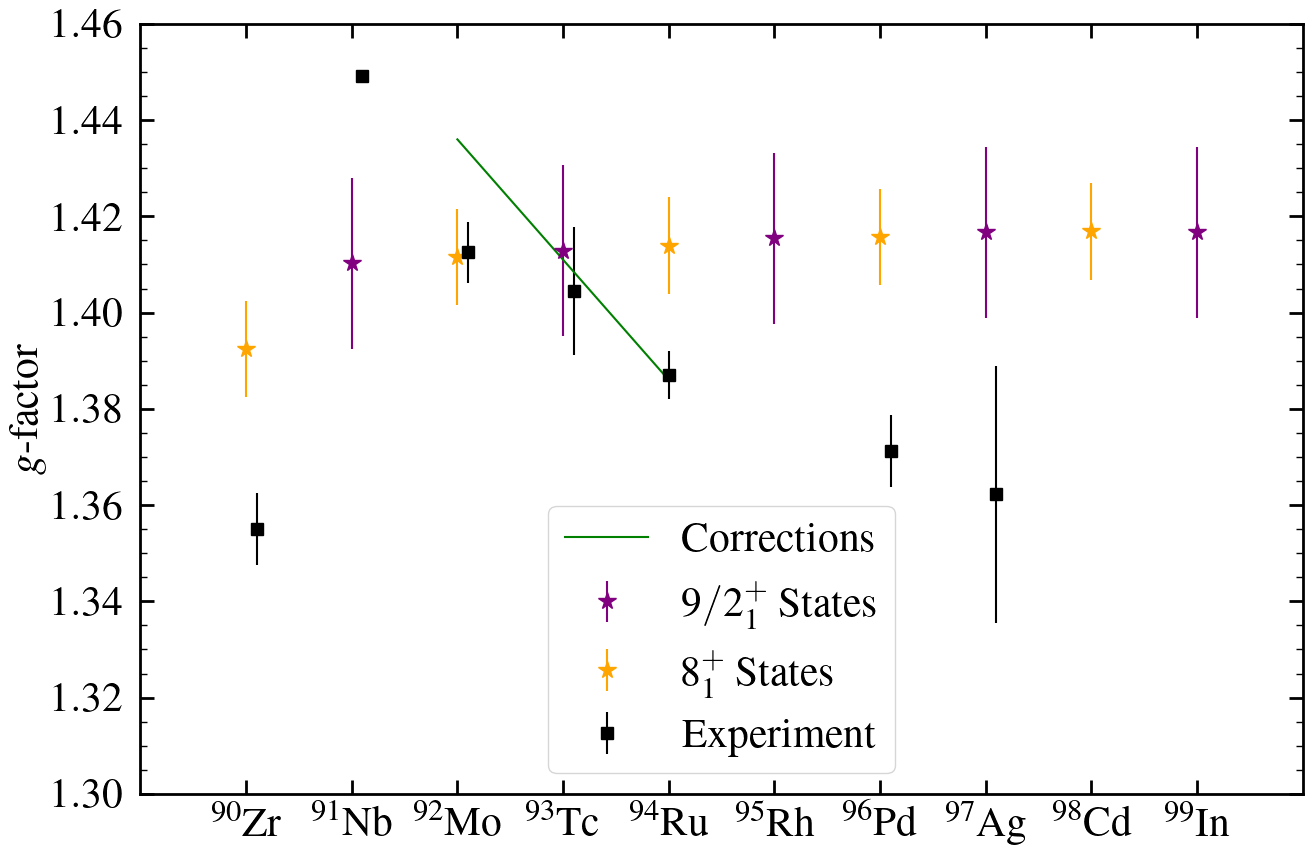}
\caption{The $g$-factors for the low-lying states of a given $J$ in each of the nuclei 
with $A \geq 90$ in our model space. Results were obtained from the p35-i3 Hamiltonian. 
The corrections (green line) arising from core-polarization effects and meson exchange 
currents were carried out in \cite{polarizationCorrections}.
}
\label{gFactors}
\end{figure}

Expanding on this, consider the $g$-factor defined by $g = \mu/(\mu_NJ)$, where $\mu$ 
is the observed magnetic dipole moment of the nucleus in a given state with total angular 
momentum $J$, and $\mu_N = 0.105 $ e$\cdot$fm is the nuclear magneton. Application of the 
projection theorem for a $n$-body wavefunction described by pure configuration states of the form $j^n$ 
coupled to total angular momentum $J$ results in $g\left(j^n, J\right) = g(j)$: 
For wavefunctions that encode pure configuration states, the $g$-factor for the nucleus 
as a whole is described by the $g$-factor of a nucleon in the single-particle orbital $j$, 
and is independent of the spin $J$ of the nucleus. Looking at Fig. $\ref{gFactors}$, one 
can see the $g$-factors for nuclei with $A \geq 90$. This figure shows the $g$-factors for the $9/2^+_1$ 
states in odd-even nuclei and the $8^+_1$ states in even-even nuclei. From the uniformity of the $g$-factors 
across alternating $J$-states, we can infer that these states are well described by $\left[0g_{9/2}\right]^n$ configurations.

For states with good seniority, the $B(E2)$ between the $  j^{n}  $ 
configurations with $\nu$=2 is given by \cite{SymmSeniority}:
\begin{equation}
    B\left(E2;j^{n},J_{i} \rightarrow J_{f}\right)  =
    \left[\frac{2j+1-2n}{2j+1-2\nu }\right]^{2} B\left(E2;j^{2},J_{i} \rightarrow J_{f}\right)
    \label{eq:E2seniority}
\end{equation}

For example, for $  j=9/2  $ and $  n=4  $ (or $  n=6  $)
\begin{equation}
    B\left(E2;j^{4},\nu =2\right) = (1/9) B\left(E2;j^{2},\nu =2\right)
    \label{eq:seniorityExample}
\end{equation}

\begin{figure}
\includegraphics[scale=0.36]{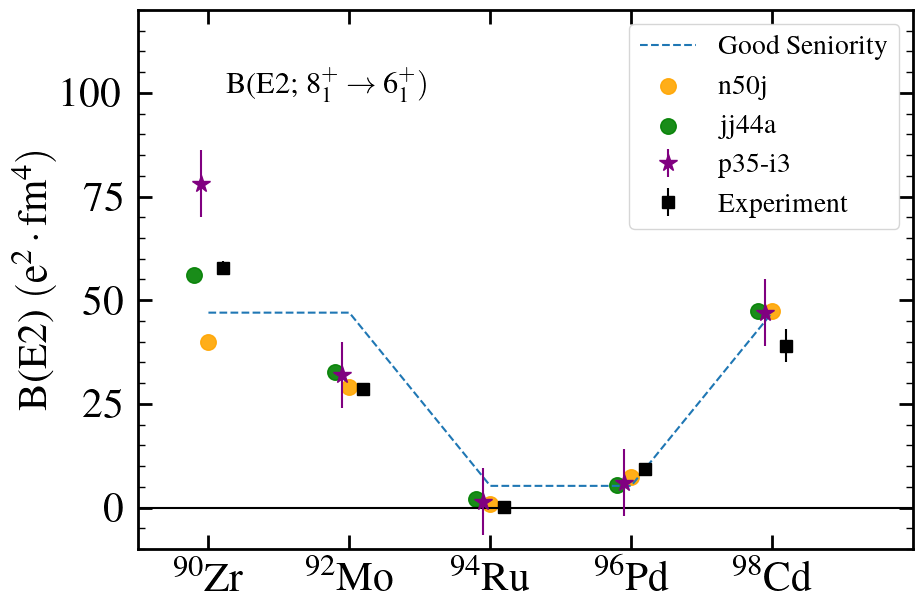}
\caption{The $B(E2)$ values for the $8^+_1 \rightarrow 6^+_1$ transition in the even-even nuclei within 
the $\pi j4$ model space, in particular the nuclei with $Z \geq 40$. The theoretical values were 
calculated using an effective charge of $e_p = 1.8$ for the $E2$ operator. 
The dashed curve indicates values calculated from Equation \ref{eq:E2seniority} obtained from \cite{SymmSeniority}.
}
\label{be86}
\end{figure}

\begin{figure}
\includegraphics[scale=0.5]{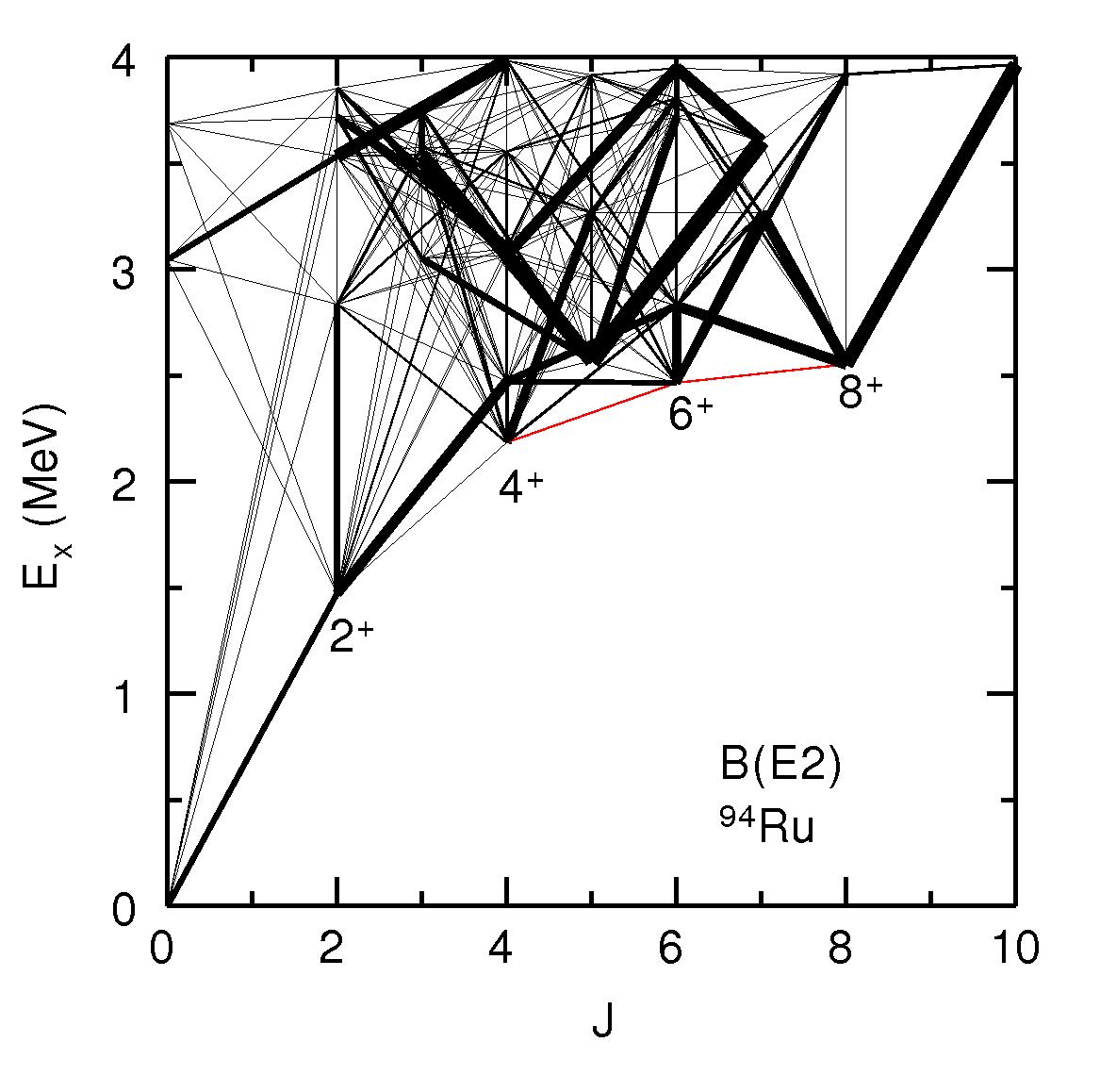}
\caption{E2-map for $^{94}$Ru. The width of the lines are proportional
to the $B(E2)$ values between the states shown. The small $B(E2)$ value 
indicated by $8_1^{+}$ to $6_1^{+}$ red line is discussed in the the text.
}
\label{ru94e2}
\end{figure} 
 
\begin{figure}
\includegraphics[scale=0.5]{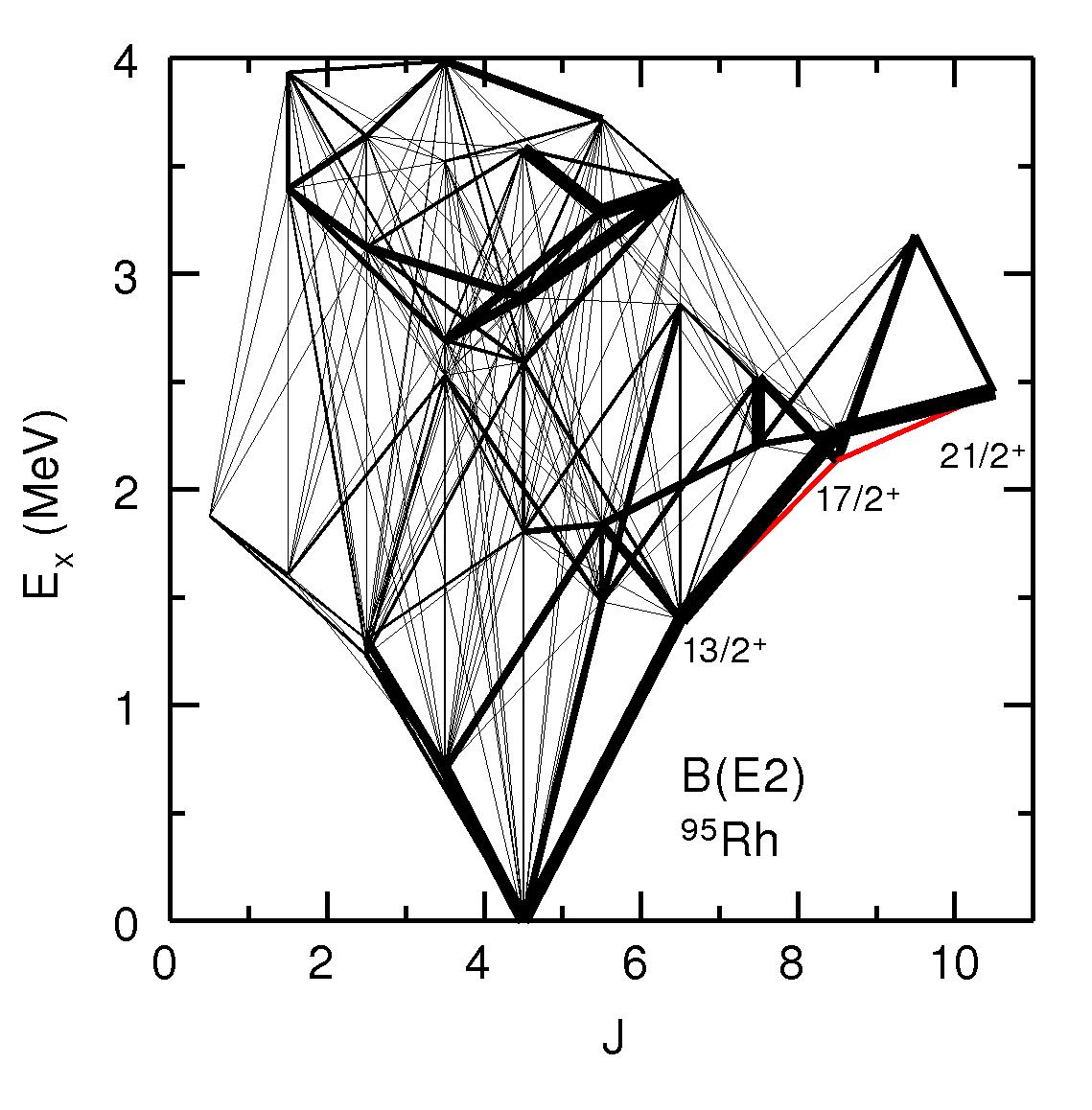}
\caption{E2-map for $^{95}$Rh. The width of the lines are proportional
to the $B(E2)$ values between the states shown. The small $B(E2)$ value 
indicated by $21/2_1^{+}$ to $17/2_1^{+}$ red line is discussed in the the text.
}
\label{rh95e2}
\end{figure} 

The $B\left(E2; 8^+_1 \rightarrow 6^+_2\right)$ for nuclei with $A \geq 90$ 
are shown in Figure \ref{be86} for the previously mentioned Hamiltonians and experiment; 
additionally there is a blue dashed curve showing the results calculated from Equation \ref{eq:E2seniority}. 
For reference, the value of $B(E2; \left[0g_{9/2}\right]^2, 8^+_1 \rightarrow 6^+_1)$ 
in this curve is calculated for the transition between states in $^{98}$Ca using the p35-i3 wavefunctions. 
The reason why $^{98}$Ca was chosen as the reference nucleus is that the wavefunctions for
the $8^+_1$ state in this nucleus has a pure $(0g_{9/2})^8$ configuration in the
$\pi-j4$ model space.
The curve can be seen to level off at $^{90}$Zr, this is because the $8^+_1$ state in 
$^{90}$Zr is also a $\left[0g_{9/2}\right]^2$ configuration; unlike the $j^n$ 
configurations in $A \geq 92$ even-even nuclei, this state involves a pair of 
protons excited out of the $1p_{1/2}$ orbit. The $8^+_1$ states even-even nuclei 
with $A \geq 92$ all involve a broken pair of protons on top of a filled $fp$ 
model space, while the broken pair in $^{90}$Zr is coupled to only the filled 
$\left\{0f_{5/2}, 1p_{3/2} \right\}$, while the $1p_{1/2}$ orbit is empty. 
One can see that both the data and model predictions agree well with the 
approximation that these wavefunctions are well described as 
$\left[0g_{9/2}\right]^n$ configurations with good seniority, 
in particular with seniority $\nu=2$.

In Table \ref{BE2table}, the $B(E2; 8^+_1 \rightarrow6^+_1)$ value for $^{94}$Ru is calculated to be 
substantially larger than the experimental value for the three Hamiltonians, 
the $B(E2; 21/2^+_1 \rightarrow 17/2^+)$ value for $^{95}$Rh, 
is calculated to be substantially smaller than experiment, and the
$B(E2)(21/2^+_1 \rightarrow 17/2^+_2)$ value is predicted to be nearly double the experimental result. 
We can trace the origin of these discrepancies to mixing of  nearby levels with 
the same spin-parity.
The E2-maps for these nuclei are shown in Figs.\ref{ru94e2} and \ref{rh95e2}.
In $^{94}$Ru there are two $6^+$ 
states within 250 keV of each other. 
The $B(E2;8^+_1 \rightarrow6^+_1)$ shown in red is
much weaker than the $B(E2; 8^+_1 \rightarrow6^+_2)$ 
shown in black immediately above the red line. In $^{95}$Rh the 
there are two $17/2^+$ states within 120 keV of each other.
In both cases B(E2) for the yrare band is much larger
that those in the yrast band. This is related to the seniority
selection rules discussed above. These $B(E2)$
can be brought into agreement with experiment with a small mixing between 
the two $6^+$ states in $^{94}$Ru and the two $17/2^+$
states in $^{95}$Rh.
The required mixing matrix element is about  25 keV. 
It is possible that this could be explained by a small readjustment
of the TBME. At this stage we ae not able to include
$B(E2)$ values in the determination of the effective TBME.
It is also possible that this is due to three-body interactions
that are not included in the calculations.


\section{Conclusion}
We discussed results for magnetic dipole moments, electric quadrupole moments, and the $B(M1)$ and $B(E2)$ for electromagnetic transitions
obtained from new effective Hamiltonians for nuclei with 50 neutrons, 
involving the $\left\{0f_{5/2}, 0f_{3/2},1p_{1/2}, 0g_{9/2}\right\}$ proton model space.
Agreement with measured values is excellent.  We list several cases which have not yet been measured, and are able to provide 
values for results for others. More data are needed towards the $^{78}$Ni and $^{100}$Sn ends of the model space.
The systematics for low-lying states as a function of $Z$ are correlated with 
the orbital occupations.


\section{Acknowledgments}

We acknowledge support from the National Science Foundation grant PHY-2110365.
\pagebreak
\bibliography{refs}

\end{document}